\documentstyle[aps,floats,epsf,twocolumn]{revtex}
 
\begin{document} \draft

\noindent
{\bf Comment on ``Confinement of Slave Particles in U(1)
Gauge Theories of Strongly Interacting Electrons" }

\vspace{10pt}

In a recent Letter \cite{nayak}, Nayak argued that slave partices
are always confined in gauge theories of strongly-correlated electron
systems like the t-J model.
The argument mostly relies on Elitzur's theorem and the fact that the 
effective gauge theory under discussion is at infinite coupling.
On the other hand, in the previous papers \cite{ichinose-matsui}
we studied dynamics of the gauge theory of the t-J model showing that
the slave particles are in the Coulomb phase {\it below} certain critical 
temperature $T_{\rm CSS}$ that depends on hole doping.
Thus, with a finite 3D coupling, the spin-charge separation occur, while
in pure 2D, slave particles interact via logarithmic potential (in 
contrast to the linear-rising confining potential).  
We comment on the discrepancy between the results in Refs.1,2.

First, the existence of the deconfining Coulomb phase {\em does not}
contradicts Elitzur's theorem which says that expectation values
of any gauge-variant quantities are vanishing.
A well-known example is the weak-coupling phase of the U(1) gauge
theory in $(3+1)$ dimensions which corresponds to the quantum 
electrodynamics. Mean field theory can also describe deconfining phases  
enforcing Elitzur's theorem \cite{drouffe}. 
Secondly, though the gauge coupling is infinite in the gauge theories of
the t-J model as Nayak argued, this fact {\em does not} necessarily
mean that the slave particles are confined.
Actually, it is known at present that the pure SU(3) gauge theory in $(3+1)$ 
dimensions is always in the confinement phase at $T=0$ regardless of strength
of the coupling.
However, if light $N_f$-flavour fermions ($N_f$ is the number of flavours)
are coupled to the SU(3) gauge theory, the system is in the 
{\em deconfinemnt phase even at infinite coupling} for $N_f>7$
\cite{iwasaki}.
This means that matter fields which couple to the gauge field strongly
influence the phase structure of the system.
Similar phenomenon happens in the gauge theory of the t-J model
as we showed in Ref.2.
Spinons and holons generate there  nontrivial gauge dynamics and the
deconfinement phase appears at low $T$.
We employed a non-perturbative method, not relying upon
the perturbative calculations contrary to Nayak's claim in Ref.1.

Let us go into details of the arguments in Ref.1.
The effective low-energy Lagrangian (8) is obtained by
simply ignoring  high-energy modes in Eq.(6).
However, in order to obtain an effective model, the high-energy (and
high-momentum) modes must be integrated out. 
In some cases, this procedure does not generate any important terms
that influence the dynamics of the low-energy modes.
However, in the present case, it is conceivable that ``kinetic terms"
of the gauge field  appear effectively from the high-energy modes.
For 1D quantum spin models, this possibility is 
realized \cite{ichinose-kayama}.
The form of the continuum Lagrangian (8) is essential in order to solve
the ``constraints" on holons and spinons.
In presence of a kinetic term of the gauge field, the constaints
are {\em not} strictly satisfied by the low-energy modes, 
though they are strictly satisfied {\em before}
integration over the high-energy modes.
To judge whether the deconfinement of the slave particles
occurs in {\em low-energy} quasiexcitations, 
careful investigation on the gauge dynamics of the low-energy theory 
that is faithful to the t-J model is necessary \cite{ichinose-matsui}.

In Ref.1, the 1D t-J model at half filling 
is dicussed for illustrative purpose.
However 1D gauge systems are quite different from those
in higher dimensions because of the lack of the
transverse direction.
Study of the t-J model on a chain and a ladder in
the framework of the slave-fermion gauge theory is 
 made in Ref.6. 
Gauge field dynamics is nontrivial and
it explains clearly 
and self consistently why quasiexciations on a chain
and a ladder are so different, reflecting that the  coefficient of the
CP$^{1}$ $\theta$-term is $\pi$ for chain whereas $2\pi$ for ladder.
 
For the 2D case, there are also problems to be solved before the conclusions
in Ref.1 are accepted.
Here again the possiblity of appearance of the gauge-field kinetic term is 
ignored.
This is the essential point to discuss the confinement problem of the
slave particles in the present context.
Furthermore, restoration of the spin SU(2) symmetry  must be shown
in a convincing way from (25). 
This is a crucial test of the effective action (25).

To summarize, (1) Matter fields which couple to gauge field can change
drastically the phase structure of the system, so study of 
the gauge dynamics is indispensable  to identify quasiexcitations
 at low enegies. (2) To obtain the low-energy effective theory, possiblity
of the gauge-field kinetic term should be addressed instead of
taking the naive continuum limit.  Appearance of the kinetic 
term prevents one from solving the constraint. 
(3) Rewriting the Lagrangian in terms of the gauge-invariant
variables does {\em not} directly implies that particles 
with nonvanishing gauge charges are confined.

  \vspace{10pt}

  \noindent
  
  Ikuo Ichinose

 Institute of Physics

  University of Tokyo

  Komaba, Tokyo, 153-8902, Japan 
  
  \vspace{10pt}

  \noindent
  
  Tetsuo Matsui
  
  Department of Physics
  
  Kinki University
  
  Higashi-Osaka, 577-8502, Japan


\begin{thebibliography}{99}
\bibitem{nayak} C.Nayak, Phys. Rev. Lett. {\bf 85},
178 (2000).
\bibitem{ichinose-matsui} I.Ichinose and T.Matsui, Nucl.Phys. {\bf B394},
281 (1993); Phys.Rev. {\bf B51}, 11860 (1995).
\bibitem{drouffe}
J.M.Drouffe, Nucl.Phys. {\bf B170},
211 (1980).
\bibitem{iwasaki}Y.Iwasaki, K.Kanaya, S.Sakai and T.Yoshie, \\
Phys.Rev.Lett. {\bf 69},
21 (1992).
\bibitem{ichinose-kayama}I.Ichinose and Y.Kayama, Nucl.Phys. {\bf B522},
569 (1998).
\bibitem{chain-ladder} I.Ichinose and T.Matsui, Phys.Rev. {\bf B57},
13790 (1998).


\end{thebibliography}
\end{document}